# Ethical Appetite: Consumer Preferences and Price Premiums for Animal Welfare-Friendly Food Products

Voraprapa Nakavachara[1]
Chanon Thongtai[1,+]
Thanarat Chalidabhongse[2]
Chanathip Pharino[3]

8 March 2025

Abstract:

This study examines how consumer attitudes toward animal welfare influence food selection and pricing using real-world market data from a Swiss supermarket. Our findings indicate that higher animal welfare standards are consistently associated with higher prices, suggesting that ethical considerations play a significant role in generating price premiums based on consumer preferences. On average, a one-point increase in the animal welfare score (ranging from 1 to 5, with 5 being the highest) corresponds to a 16.4% price increase, with the effect being most pronounced in Dairy & Eggs (25.3%), compared to Meat & Fish (14.3%). These results highlight the psychological and behavioral factors underlying consumer preferences for ethically produced foods. Additionally, we find limited evidence of a price premium for climate-friendly food products, observed only in Yogurts & Desserts, a subcategory within Dairy & Eggs. Our findings contribute to the understanding of how ethical food attributes influence consumer decision-making and pricing in retail settings.

[1] Faculty of Economics, Chulalongkorn University.
[2] Department of Computer Engineering, Faculty of Engineering, Chulalongkorn University.
[3] Department of Environmental Engineering, Faculty of Engineering, Chulalongkorn University.
[+] Corresponding Author's Email: Chanon.Th@chula.ac.th.

This Quick Win research project is supported by Ratchadapiseksompotch Fund Chulalongkorn University.

# 1. Introduction

In recent years, consumer food choices have expanded beyond taste, price, and quality to include ethical considerations, particularly animal welfare and climate impact. As income levels rise, consumers, especially in high-income European markets, increasingly prioritize ethical production methods (Cembalo et al., 2016). The growing demand for animal welfare-friendly products reflects a broader shift in consumer behavior, shaped by psychological, social, and cultural factors, as animal welfare is not solely assessed through biological indicators but is also influenced by societal values (Ohl et al., 2012). In addition, consumers who prioritize ethical food attributes often perceive these products as higher quality, leading to a willingness to pay a premium (Zander & Hamm, 2010; Cornish et al., 2020). Despite the growing interest in ethical food choices, empirical evidence on the price premium associated with these attributes remains limited.

While numerous studies have examined consumer willingness to pay for animal welfare-friendly foods through survey-based methods and experiments, less is known about how these preferences translate into actual price differences in retail markets. Existing literature suggests that moral and psychological factors influence ethical food choices, yet empirical evidence on how these factors manifest in retail pricing and consumer behavior remains limited. This study contributes to the literature by quantifying the price premium for animal welfare-friendly food products in a real-world supermarket setting, offering insights into how consumer food attitudes translate into economic outcomes.

Using real market data from a Swiss supermarket, this study examines whether higher animal welfare scores correspond to a price premium. Our findings indicate that higher animal welfare standards are consistently associated with higher prices, suggesting that consumers are willing to pay more for ethical food attributes. The premium is most pronounced in Dairy & Eggs category compared to Meat & Fish, indicating that ethical preferences may vary by food type. Additionally, we find limited evidence of a price premium for climate-friendly products, observed only in the Yogurts & Desserts subcategory within Dairy & Eggs.

# 2. Literature Review and Research Framework

This section explores multiple theoretical perspectives, including consumer preference theory, signaling theory, and nudging, to examine how and why consumers choose animal welfare-friendly food products and to understand their willingness to pay a price premium.

## 2.1 Consumer Preference for Ethical Food Choices

Consumer preference theory suggests that individuals make decisions based on both tangible and intangible attributes to maximize their utility, with willingness to pay commonly used to assess preferences (Techa-Erawan et al., 2024). Preferences in economic terms can be classified as transitive and non-transitive. Transitive preferences imply that if a consumer prefers product A over B and B over C, they should also prefer A over C (Shafer, 1990). Research suggests that food products are more likely to follow transitive preferences due to repeated consumption and sensory feedback (Guadalupe-Lanas et al., 2020). However, ethical considerations, such as climate concerns, may introduce non-transitive preferences, influenced by complex social norms and cultural values (Andre et al., 2021).

Consumers who prioritize ethical food attributes often perceive these products as higher quality, leading to a willingness to pay a premium (Zander & Hamm, 2010; Cornish et al., 2020). Cembalo et al. (2016) find that, in five European countries, self-transcendence values strongly influence attitudes toward animal welfare-friendly foods, while Miranda-de la Lama et al. (2019) show that consumers in Mexico are willing to pay more for certified welfare-friendly products but seek clearer regulations and information.

## 2.2 Labeling as Information Signaling in Food Selection

Signaling (Spence, 1978; 2002) plays a crucial role in communicating specific product attributes to decision-makers. In modern trade, any labels on a product's packaging serve as signals in practice. Labels and certificates for animal welfare-friendly food products act as indicators of ethical production methods, influencing consumer perception. A study by Araya et al. (2022) examining the effects of nutritional warning labels in Chile found that cereals, chocolates, and cookies with labels experienced price increases, whereas unlabeled items either had price reductions or negligible changes. This suggests that price-sensitive consumers are more likely to prefer unlabeled products, prompting companies to adjust prices in response to consumer reactions to labeling (Pachali et al., 2023). Additionally, country-of-origin labeling influences consumers' willingness to pay, as they tend to prefer products from their favored regions (Loureiro & Umberger, 2003).

Research on animal welfare labeling highlights its role in product differentiation and increased willingness to pay (Napolitano et al., 2010). Toma et al. (2012) find that access to animal welfare information and labeling significantly influences consumers' willingness to switch shopping locations, highlighting the need for more effective labeling systems. Gorton et al. (2023) investigate willingness to pay for chicken meat with and without animal welfare labels, revealing that consumers are willing to pay more for labeled products, though emotional appeals may influence decisions more than the actual welfare standards. Similarly, Cornish et al. (2020) find that detailed welfare information on labels increases consumer trust and purchase likelihood, particularly among those with higher animal empathy.

## 2.3 Behavioral Nudges in Ethical Food Choices

Behavioral nudging (Thaler & Sunstein; 2021) involves subtly influencing consumer decisions without restricting choices or altering economic incentives. Nudges work by changing the way options are presented, leveraging psychological and cognitive biases to guide individuals toward desired behaviors. Behavioral nudging techniques can influence consumer purchasing patterns for animal welfare-friendly products. Van der Vliet (2024) tests the impact of nudging plant-based alternatives over meat and dairy in an online supermarket experiment but finds no significant effect, suggesting that additional strategies are needed to shift consumer behavior toward plant-based proteins. Conversely, Weingarten et al. (2024) test a visibility nudge in a virtual supermarket, showing that placing animal welfare-labeled products in prominent shelf locations significantly increases purchases, regardless of price sensitivity.

## 2.4 Hypothesis Development

Based on consumer preference theory and signaling theory, we anticipate that consumers prefer ethical food products (i.e., animal welfare-friendly and/or climate-friendly foods) and are willing to pay a premium for them. These preferences are expected to be reflected in actual price differences in retail markets. Therefore, our hypotheses are as follows:

H1: Food products with higher animal welfare scores are associated with higher prices.

H2: Food products with better climate impact scores are associated with higher prices.

## 3. Data and Methodology

### 3.1 Data

We utilize data from a Swiss supermarket (as of February 2025), focusing on food products in the Dairy & Eggs and Meat & Fish categories, where animal welfare is a relevant factor. We include only products with an assigned animal welfare score, which ranges from 1 to 5, with 5 indicating the highest standard. The animal welfare score assessment evaluates key aspects of the production process, including but not limited to sick animal medication, regular exercise, mother-bonded rearing, and the use of painful procedures. Additionally, most of these products are assigned a climate impact score, which is determined by their carbon dioxide equivalent emissions per kilogram of product (kg $CO_2$eq/kg). The climate impact score also ranges from 1 to 5, with 5 representing the most favorable rating. To ensure consistency in our analysis, we retain only products for which the size variable is specified in grams. Products measured in volume (e.g., liters) are excluded, and sizes originally recorded in kilograms are converted to grams. The final dataset comprises 550 products, with 264 in the Dairy & Eggs category and 286 in the Meat & Fish category.

Table 1 presents the summary statistics of the dataset. The price of the products ranges from 0.90 CHF to 38.65 CHF, with an average price of 6.53 CHF. The product size varies between 50 grams and 1,325 grams, with an average of 329.9 grams. Additional nutritional attributes are also summarized in the table. On average, the products contain 15.93 grams of carbohydrates, 43.29 grams of fat, 50.56 grams of protein, 1.40 grams of fiber, 3.24 grams of salt, and provide 657.54 kcal of energy. Regarding sustainability and ethical considerations, the average animal welfare score is 3.25, while the average climate impact score is 2.30. Both scores range from 1 to 5, with higher values indicating better performance in these aspects.

Tables 2 and 3 summarize the statistics for the Dairy & Eggs and Meat & Fish categories, respectively. The Dairy & Eggs category is further divided into three subcategories: Cheese, Fresh Food, Milk, Butter & Eggs, and Yogurts & Desserts. Meanwhile, the Meat & Fish category consists of two subcategories: Cold Cuts and Meat & Poultry.

The average price of products in the Dairy & Eggs category is 4.32 CHF, with an average size of 1.36 grams. In terms of nutritional composition, these products contain, on average, 28.32 grams of carbohydrates, 51.58 grams of fat, 40.69 grams of protein, 1.53 grams of fiber, 2.56 grams of salt, and provide 744.02 kcal of energy. Concerning sustainability and ethical considerations, the average animal welfare score is 3.20, while the average climate impact score is 2.84.

On the other hand, products in the Meat & Fish category have a higher average price of 8.57 CHF and a larger average size of 3.49 grams. Their nutritional profile differs significantly from the Dairy & Eggs category, with an average of 4.55 grams of carbohydrates, 35.67 grams of fat, 59.62 grams of protein, 1.27 grams of fiber, 3.86 grams of salt, and 578.12 kcal of energy. Regarding sustainability indicators, the average animal welfare score is slightly higher at 3.30, whereas the average climate impact score is lower at 1.79.

## 3.2 Methodology

To test our hypothesis, we examine whether ethical attributes—such as a higher animal welfare score and a better climate impact score—are associated with a price premium, potentially reflecting consumers' willingness to pay for ethical consumption. We employ the following empirical model:

$$y_i = \beta_0 + \beta_1 a_i + \beta_2 c_i + \boldsymbol{\beta'_3 x_i} + \theta_{country} + \gamma_{subtype} + \varepsilon_i \qquad (1)$$

where $y_i$ represents the natural logarithm of the price of product i. The variable $a_i$ and $c_i$ correspond to product i's animal welfare score and climate impact score, respectively. The vector $\boldsymbol{x_i}$ includes product attributes such as size and nutritional components (carbohydrates, fat, protein, fiber, salt, and energy). $\beta_0$ is the intercept term. $\beta_1$ and $\beta_2$ are coefficients of the animal welfare score and climate impact score, respectively. These are the coefficients of our interest. $\boldsymbol{\beta'_3}$ is a vector of coefficients of product attributes. Fixed effects for country of origin ($\theta_{country}$) and product subtype ($\gamma_{subtype}$) are included to account for unobserved heterogeneity. The error terms $\varepsilon_i$ are clustered at the subtype level to account for within-group correlations in pricing.

Subtypes represent a more granular classification than subcategories. For instance, within the Cheese subcategory, subtypes include fresh cheese, sliced cheese, and hard & semi-hard cheese, among others. Similarly, the Fresh Food, Milk, Butter & Eggs subcategory comprises fresh pasta, milk, margarine, eggs, and additional subtypes. The Yogurts & Desserts subcategory includes yogurts, children's desserts, and other similar products. In the Cold Cuts subcategory, examples of subtypes include bacon, smoked ham, and salami, while the Meat & Poultry subcategory encompasses beef, chicken, rabbit, pork, and several other meat varieties.

## 4. Results and Discussion

Table 4 presents the main regression results. Columns (1) and (2) report estimates based on the full sample, while Columns (3) and (4) focus on the Dairy & Eggs category, and Columns (5) and (6) examine the Meat & Fish category. The odd-numbered columns (1, 3, and 5) show results from models that include the animal welfare score but exclude the climate impact score, whereas the even-numbered columns (2, 4, and 6) report results from models incorporating both ethical attributes.

Across all specifications, the coefficients on the animal welfare score are positive and statistically significant, indicating a price premium for animal welfare-friendly products. On average, a one-point increase in the animal welfare score is associated with a 16.4% increase in price. This premium is more pronounced in the Dairy & Eggs category compared to the Meat & Fish category (25.3% vs. 14.3%). In contrast, the coefficients on the climate impact score are not statistically significant in any of the models, suggesting that climate-friendly products do not command a price premium.

The impact of nutritional components—such as carbohydrates, fat, protein, fiber, salt, and energy—on product prices appears to be inconclusive. For instance, higher protein content is associated with higher prices in the full-sample models (Columns 1 and 2) and the Dairy & Eggs category (Columns 3 and 4). However, in the Meat & Fish category, the effect of protein is hardly statistically significant or even negative, suggesting that its influence on pricing varies across product categories.

Table 5 presents the regression results for Dairy & Eggs subgroups. Columns (1) and (2) report estimates based on the Cheese subgroup, while Columns (3) and (4) focus on the Fresh Food, Milk, Butter & Eggs subgroup, and Columns (5) and (6) examine the Yogurts & Desserts subgroup. Consistently across all specifications, the coefficients on the animal welfare score remain positive and statistically significant, reinforcing the presence of a price premium for animal welfare-friendly products. Specifically, a one-point increase in the animal welfare score is associated with a 22.9%, 22.4%, and 31.5% increase in price for the Cheese, Fresh Food, Milk, Butter & Eggs, and Yogurts & Desserts subcategories, respectively.

Regarding climate impact, the coefficient is positive and statistically significant only for the Yogurts & Desserts subcategory. Thus, we find only limited evidence of a price premium for climate-friendly products. The relationship between nutritional components—such as carbohydrates, fat, protein, fiber, salt, and energy—and product prices appears inconsistent, with significance and effects differing across categories. This variability suggests that nutritional attributes do not uniformly influence pricing.

Table 6 presents the regression results for the Meat & Fish subcategories. Columns (1) and (2) show estimates for the Cold Cuts subgroup, while Columns (3) and (4) pertain to the Meat & Poultry subgroup. Consistently across all models, the animal welfare score exhibits a positive and statistically significant relationship with price. Specifically, each additional point in the animal welfare score corresponds to a 16.2% increase in price for Cold Cuts and a 17.4% increase for Meat & Poultry.

Conversely, the climate impact score does not display statistical significance in any model, indicating that climate-friendly attributes do not appear to influence pricing. The effect of nutritional factors—such as carbohydrates, fat, protein, fiber, salt, and energy—on price remains ambiguous, with variations observed across different product categories.

The regression results provide consistent evidence that animal welfare-friendly products command a price premium across all categories and subcategories. The animal welfare score is positively and significantly associated with price, indicating that consumers are willing to pay more for products with higher welfare standards. Notably, the premium appears more pronounced in the Dairy & Eggs category compared to Meat & Fish.

In contrast, we find little evidence of a price premium for climate-friendly products. The climate impact score is generally not statistically significant, except for the Yogurts & Desserts subcategory, suggesting that consumer willingness to pay for climate impact attributes is limited.

The impact of nutritional components on pricing remains inconclusive, as their effects vary across product categories and subcategories. While some factors, such as protein content, are associated with higher prices in certain models, these relationships are inconsistent.

## 5. Conclusion

This study contributes to the growing body of research on consumer food selection behavior by examining the market premium for animal welfare-friendly food products using real-world price data from a Swiss supermarket. Our findings indicate that higher animal welfare standards are consistently associated with higher prices, suggesting that consumers are willing to pay a premium for ethical attributes in food products. Specifically, a one-point increase in the animal welfare score is linked to a 16.4% price increase, with the effect being most pronounced in the Dairy & Eggs category (25.3%), compared to Meat & Fish (14.3%).

These results highlight the psychological and behavioral factors influencing food selection, particularly the role of ethical considerations in shaping willingness to pay, which translates into a price premium in the retail market. The stronger price premium observed in dairy and egg products suggests that ethical food preferences may vary by product type.

Interestingly, our findings show limited evidence of a price premium for climate-friendly food attributes, which appears only in the Yogurts & Desserts subcategory within Dairy & Eggs. This suggests that while animal welfare is a key driver of ethical food pricing, climate considerations may not yet command the same consumer-driven price differentiation, at least in the Swiss supermarket context.

Future research could further explore the sensory and psychological dimensions of ethical food selection, investigating how taste expectations, cultural influences, and social norms shape consumer willingness to pay for ethical attributes. Additionally, studies across different markets and income groups could provide a broader understanding of how ethical food premiums vary across regions and demographic segments.

Overall, our findings reinforce the growing role of ethical considerations in consumer food choices, demonstrating how animal welfare attributes influence real-world food pricing and selection decisions. As consumer preferences continue to evolve, understanding the behavioral and cultural drivers of ethical food selection will remain critical for shaping future food markets and sustainability policies.

**Declaration of generative AI and AI-assisted technologies in the writing process**

During the preparation of this work the author(s) used GPT-4 in the writing process to improve the readability and language of the manuscript. After using this tool/service, the author(s) reviewed and edited the content as needed and take(s) full responsibility for the content of the published article.

Table 1: Summary Statistics (All Observations)

| Variable | Obs | Mean | Std. dev. | Min | Max |
|---|---|---|---|---|---|
| Price (CHF) | 550 | 6.53 | 5.74 | 0.90 | 38.65 |
| LnPrice | 550 | 1.57 | 0.78 | -0.11 | 3.65 |
| Size (g) | 550 | 329.90 | 213.66 | 50.00 | 1325.00 |
| Carb (g) | 541 | 15.93 | 29.79 | 0.00 | 218.35 |
| Fat (g) | 541 | 43.29 | 46.95 | 0.00 | 350.00 |
| Protein (g) | 541 | 50.56 | 44.86 | 0.00 | 250.00 |
| Fibre (g) | 514 | 1.40 | 2.66 | 0.00 | 27.75 |
| Salt (g) | 541 | 3.24 | 3.93 | 0.00 | 24.15 |
| Energy (kcal) | 541 | 657.54 | 523.24 | 92.00 | 3631.25 |
| Animal_Welfare | 550 | 3.25 | 0.69 | 1.00 | 5.00 |
| Climate_Impact | 514 | 2.30 | 0.91 | 1.00 | 4.00 |

| Category | Freq. | Percent | Cum. |
|---|---|---|---|
| Dairy & Eggs | 264 | 48.00 | 48.00 |
| Meat & Fish | 286 | 52.00 | 100.00 |
| Total | 550 | 100.00 | |

Table 2: Summary Statistics (Dairy & Eggs)

| Variable | Obs | Mean | Std. dev. | Min | Max |
|---|---|---|---|---|---|
| Price (CHF) | 264 | 4.32 | 4.07 | 0.90 | 23.50 |
| LnPrice | 264 | 1.17 | 0.72 | -0.11 | 3.16 |
| Size (g) | 259 | 1.36 | 1.10 | 0.17 | 7.80 |
| Carb (g) | 259 | 28.32 | 37.63 | 0.00 | 218.35 |
| Fat (g) | 259 | 51.58 | 57.19 | 0.00 | 350.00 |
| Protein (g) | 259 | 40.69 | 42.92 | 0.50 | 250.00 |
| Fibre (g) | 258 | 1.53 | 3.51 | 0.00 | 27.75 |
| Salt (g) | 259 | 2.56 | 3.44 | 0.00 | 20.00 |
| Energy (kcal) | 259 | 744.02 | 608.40 | 103.70 | 3631.25 |
| Animal_Welfare | 264 | 3.20 | 0.52 | 2.00 | 4.00 |
| Climate_Impact | 249 | 2.84 | 0.90 | 1.00 | 4.00 |

| Type | Freq. | Percent | Cum. |
|---|---|---|---|
| Cheese | 113 | 42.8 | 42.8 |
| Fresh Food, Milk, Butter & Eggs | 41 | 15.53 | 58.33 |
| Yogurts & Desserts | 110 | 41.67 | 100 |
| Total | 264 | 100 | |

Table 3: Summary Statistics (Meat & Fish)

| Variable | Obs | Mean | Std. dev. | Min | Max |
|---|---|---|---|---|---|
| Price (CHF) | 286 | 8.57 | 6.28 | 1.75 | 38.65 |
| LnPrice | 286 | 1.94 | 0.63 | 0.56 | 3.65 |
| Size (g) | 272 | 3.49 | 2.50 | 0.59 | 15.95 |
| Carb (g) | 282 | 4.55 | 11.57 | 0.00 | 120.00 |
| Fat (g) | 282 | 35.67 | 33.33 | 1.00 | 210.00 |
| Protein (g) | 282 | 59.62 | 44.75 | 0.00 | 240.00 |
| Fibre (g) | 256 | 1.27 | 1.32 | 0.00 | 8.05 |
| Salt (g) | 282 | 3.86 | 4.24 | 0.00 | 24.15 |
| Energy (kcal) | 282 | 578.12 | 415.98 | 92.00 | 2422.00 |
| Animal_Welfare | 286 | 3.30 | 0.80 | 1.00 | 5.00 |
| Climate_Impact | 265 | 1.79 | 0.57 | 1.00 | 3.00 |

| Type | Freq. | Percent | Cum. |
|---|---|---|---|
| Cold Cuts | 151 | 52.8 | 52.8 |
| Meat & Poultry | 135 | 47.2 | 100 |
| Total | 286 | 100 | |

Table 4: Regression Results (Main)

| VARIABLES | (1) LnPrice | (2) LnPrice | (3) LnPrice | (4) LnPrice | (5) LnPrice | (6) LnPrice |
|---|---|---|---|---|---|---|
| Size | 0.000358 | 0.000350 | 0.000635*** | 0.000639*** | -0.000544 | -0.000596 |
|  | (0.000257) | (0.000241) | (0.000206) | (0.000181) | (0.000597) | (0.000659) |
| Carb | 0.00591*** | 0.00331* | 0.00254* | 0.00172 | -0.0925** | -0.0896** |
|  | (0.00195) | (0.00191) | (0.00138) | (0.00132) | (0.0375) | (0.0353) |
| Fat | 0.0146*** | 0.00918* | 0.0122*** | 0.0106*** | -0.205** | -0.195** |
|  | (0.00531) | (0.00527) | (0.00301) | (0.00249) | (0.0844) | (0.0844) |
| Protein | 0.0108*** | 0.00824*** | 0.00864*** | 0.00684*** | -0.0794* | -0.0750* |
|  | (0.00198) | (0.00211) | (0.00138) | (0.00164) | (0.0373) | (0.0369) |
| Fibre | -0.0245* | -0.0280** | -0.00674 | -0.0102** | -0.0373 | -0.0305 |
|  | (0.0136) | (0.0137) | (0.00608) | (0.00405) | (0.0220) | (0.0204) |
| Salt | 0.0297** | 0.0314** | -0.0279 | -0.0127 | 0.0493*** | 0.0517*** |
|  | (0.0128) | (0.0127) | (0.0188) | (0.0190) | (0.0114) | (0.0126) |
| Energy | -0.00158** | -0.000976* | -0.00106*** | -0.000873*** | 0.0224** | 0.0213** |
|  | (0.000583) | (0.000574) | (0.000335) | (0.000284) | (0.00944) | (0.00943) |
| Animal_Welfare | 0.160*** | 0.164*** | 0.242*** | 0.253*** | 0.147*** | 0.143*** |
|  | (0.0307) | (0.0312) | (0.0447) | (0.0412) | (0.0262) | (0.0265) |
| Climate_Impact |  | -0.0208 |  | -0.0517 |  | -0.0528 |
|  |  | (0.0424) |  | (0.0852) |  | (0.0560) |
| Constant | 0.838*** | 0.871*** | 0.415* | 0.402* | 0.0235 | 0.107 |
|  | (0.195) | (0.219) | (0.209) | (0.214) | (0.196) | (0.235) |
|  |  |  |  |  |  |  |
| Observations | 514 | 487 | 258 | 245 | 256 | 242 |
| R-squared | 0.865 | 0.866 | 0.888 | 0.882 | 0.805 | 0.808 |
| Country Dummy | Yes | Yes | Yes | Yes | Yes | Yes |
| Category Dummy | subtypeID | subtypeID | subtypeID | subtypeID | subtypeID | subtypeID |
| Cluster | subtypeID | subtypeID | subtypeID | subtypeID | subtypeID | subtypeID |
| Obs. Group | All | All | Dairy & Eggs | Dairy & Eggs | Meat & Fish | Meat & Fish |

Robust standard errors in parentheses
*** p<0.01, ** p<0.05, * p<0.1

Table 5: Regression Results (Dairy & Eggs Subgroups)

| VARIABLES | (1) LnPrice | (2) LnPrice | (3) LnPrice | (4) LnPrice | (5) LnPrice | (6) LnPrice |
|---|---|---|---|---|---|---|
| Size | 0.00165 | 0.00209** | -0.000703 | -0.000609 | -0.00150** | -0.00158*** |
|  | (0.000943) | (0.000712) | (0.000790) | (0.000939) | (0.000336) | (0.000305) |
| Carb | 0.00464 | 0.00765 | -0.0733** | -0.0583 | -0.00769 | -0.0116 |
|  | (0.00932) | (0.00670) | (0.0273) | (0.0477) | (0.0114) | (0.0150) |
| Fat | 0.0215 | 0.0262** | -0.146** | -0.114 | -0.00982 | -0.0183 |
|  | (0.0119) | (0.00830) | (0.0597) | (0.105) | (0.0265) | (0.0340) |
| Protein | 0.00953* | 0.00862* | -0.0408 | -0.0276 | 0.0123 | 0.0101 |
|  | (0.00490) | (0.00399) | (0.0393) | (0.0533) | (0.0112) | (0.0143) |
| Fibre | -0.00976 | -0.0762 | -0.0157 | -0.0104 | -0.00267 | -0.00398 |
|  | (0.0620) | (0.0447) | (0.0141) | (0.0220) | (0.00534) | (0.00473) |
| Salt | -0.0331 | -0.00790 | 0.0372 | 0.0342 | 0.625 | 0.636* |
|  | (0.0216) | (0.0138) | (0.0529) | (0.0556) | (0.297) | (0.288) |
| Energy | -0.00216 | -0.00273** | 0.0168** | 0.0132 | 0.00247 | 0.00348 |
|  | (0.00144) | (0.00106) | (0.00680) | (0.0118) | (0.00267) | (0.00352) |
| Animal_Welfare | 0.215** | 0.229** | 0.235*** | 0.224* | 0.323*** | 0.315*** |
|  | (0.0636) | (0.0648) | (0.0551) | (0.101) | (0.0260) | (0.0250) |
| Climate_Impact |  | -0.104 |  | -0.0593 |  | 0.249*** |
|  |  | (0.0560) |  | (0.115) |  | (0.0370) |
| Constant | -0.636*** | -0.414* | 0.411 | 0.453 | -0.958** | -1.733*** |
|  | (0.171) | (0.210) | (0.314) | (0.514) | (0.247) | (0.243) |
| Observations | 108 | 101 | 41 | 40 | 109 | 104 |
| R-squared | 0.913 | 0.919 | 0.725 | 0.719 | 0.698 | 0.711 |
| Country Dummy | Yes | Yes | Yes | Yes | Yes | Yes |
| Category Dummy | subtypeID | subtypeID | subtypeID | subtypeID | subtypeID | subtypeID |
| Cluster | subtypeID | subtypeID | subtypeID | subtypeID | subtypeID | subtypeID |
| Obs. Group | Dairy & Eggs Cheese | Dairy & Eggs Cheese | Dairy & Eggs Fresh & Milk & Eggs | Dairy & Eggs Fresh & Milk & Eggs | Dairy & Eggs Yogurt & Desserts | Dairy & Eggs Yogurt & Desserts |

Robust standard errors in parentheses
*** p<0.01, ** p<0.05, * p<0.1

Table 6: Regression Results (Meat & Fish Subgroups)

| VARIABLES | (1) LnPrice | (2) LnPrice | (3) LnPrice | (4) LnPrice |
|---|---|---|---|---|
| Size | -0.00194* | -0.00238* | -2.12e-05 | -8.91e-05 |
|  | (0.000876) | (0.00119) | (0.000436) | (0.000455) |
| Carb | -0.0566 | 0.00625 | -0.0889* | -0.0919* |
|  | (0.117) | (0.109) | (0.0454) | (0.0393) |
| Fat | -0.241 | -0.103 | -0.194* | -0.197* |
|  | (0.251) | (0.230) | (0.0998) | (0.0908) |
| Protein | -0.0777 | -0.0144 | -0.0783 | -0.0797* |
|  | (0.111) | (0.101) | (0.0447) | (0.0403) |
| Fibre | -0.0259 | -0.00248 | -0.0683*** | -0.0691*** |
|  | (0.0496) | (0.0479) | (0.0129) | (0.0133) |
| Salt | -0.0739* | -0.0762* | 0.0556*** | 0.0579*** |
|  | (0.0319) | (0.0324) | (0.0142) | (0.0132) |
| Energy | 0.0268 | 0.0115 | 0.0214 | 0.0218* |
|  | (0.0280) | (0.0256) | (0.0112) | (0.0102) |
| Animal_Welfare | 0.161*** | 0.162*** | 0.173** | 0.174** |
|  | (0.0261) | (0.0205) | (0.0515) | (0.0582) |
| Climate_Impact |  | -0.0740 |  | 0.00966 |
|  |  | (0.106) |  | (0.0348) |
| Constant | 0.345** | 0.349 | 0.959** | 0.942** |
|  | (0.128) | (0.319) | (0.266) | (0.302) |
|  |  |  |  |  |
| Observations | 136 | 129 | 120 | 113 |
| R-squared | 0.774 | 0.775 | 0.715 | 0.706 |
| Country Dummy | Yes | Yes | Yes | Yes |
| Category Dummy | subtypeID | subtypeID | subtypeID | subtypeID |
| Cluster | subtypeID | subtypeID | subtypeID | subtypeID |
| Obs. Group | Meat & Fish Cold Cuts | Meat & Fish Cold Cuts | Meat & Fish Meat & Poultry | Meat & Fish Meat & Poultry |

Robust standard errors in parentheses
*** p<0.01, ** p<0.05, * p<0.1